\documentclass[reprint,superscriptaddress,nofootinbib,amsmath,amssymb,aps]{revtex4-1}

\usepackage{graphicx}% Include figure files
\usepackage{dcolumn}% Align table columns on decimal point
\usepackage{bm}% bold math
\usepackage{color}
\usepackage{mathrsfs}
\usepackage[dvipsnames]{xcolor}

\def\prd{Phys. Rev. D}

\begin{document}

\preprint{APS/123-QED}

\title{Mono- and oligochromatic extreme-mass ratio inspirals} 

\author{Pau Amaro Seoane}
\affiliation{Universitat Politècnica de València, Spain}
\affiliation{Max Planck Institute for Extraterrestrial Physics, Garching, Germany}
\affiliation{Higgs Centre for Theoretical Physics, Edinburgh, UK}
\author{Yiren Lin}
\affiliation{Astronomy Department, School of Physics, Peking University, Beijing 100871, China}
\author{Kostas Tzanavaris}
\affiliation{Higgs Centre for Theoretical Physics, Edinburgh, UK}
\date{\today}

\begin{abstract}
The gravitational capture of a stellar-mass object by a supermassive black hole
represents a unique probe of warped spacetime.  The small object, typically a
stellar-mass black hole, describes a very large number of cycles before
crossing the event horizon. Because of the mass difference, we call these
captures extreme-mass ratio inspirals (EMRIs).
Their merger event rate at the Galactic Centre is negligible, but the amount of
time spent in the early inspiral is not. Early EMRIs (E-EMRIs) spend hundreds
of thousands of years in band during this phase. At very early stages, the peak
of the frequency will not change during an observational time. At later stages
in the evolution, it will change a bit and finally the EMRI explores a wide
range of them when it close to merger. We distinguish between ``monocromatic''
E-EMRIs, which do not change their (peak) frequency, oligochromatic E-EMRIs,
which explore a short range and polychromatic ones, the EMRIs which have been
discussed so far in the literature.
We derive the number of E-EMRIs at the Galactic Centre, and we also calculate 
their signal-to-noise ratios (SNR) and perform a study of parameter extraction.
We show that parameters such as the spin and the mass can be extracted with an
error which can be as small as $10^{-11}$ and $10^{-5}\,M_{\odot}$.
There are between hundreds and thousands of E-EMRIs in their monochromatic
stage at the GC, and tens in their oligochromatic phase. The SNR ranges from a
minimum of $10$ (larger likelihood) to a maximum of $10^6$ (smaller
likelihood). 
Moreover, we derive the contribution signal corresponding to the incoherent sum
of continuous sources of oligochromatic E-EMRIs with two representatives
masses; $10\,M_{\odot}$ and $40\,M_{\odot}$ and show that their curves will
cover a significant part of LISA's sensitivity curve. Depending on their level
of circularisation, they might be detected as individual sources or form a
foreground population.

\end{abstract}

                             % Classification Scheme.
                              %display desired
\maketitle

\section{Introduction}
\label{sec.intro}

When a stellar-mass compact object such as a white dwarf, a neutron star or,
more likely due to mass segregation, a black hole forms a binary with a
supermassive one (MBH), the system radiates energy in gravitational waves
\citep{Einstein1916,Einstein1918}. The large mass-ratio of the system is large,
ranging between $10^6$ for a neutron star or white dwarf forming a binary with
a MBH of mass $10^6\,M_{\odot}$, to $10^5$ for a stellar-mass blach hole of
$10\,M_{\odot}$. Therefore we call these systems extreme-mass ratio inspirals
\citep{Amaro-SeoaneLRR2012,AmaroSeoane2022,Amaro-SeoaneEtAl07}, and they
constitute one of the main targets of the Laser Interferometer Space Antenna
\citep[LISA][]{Amaro-SeoaneEtAl2017}. We can envisage these sources as a kind
of camera taking snapshots of warped spacetime around a MBH.  They will allow
us to probe very closely the event horizon of the MBH and the geometry of
spacetime in this regime of strong gravity
\citep{Amaro-SeoaneGairPoundHughesSopuerta2015}.

The timescale associated to a gravitational-wave source can be approximated
thanks to the work of \cite{Peters64}. This timescale has the strongest
dependency with the semimajor axis of the binary, which is to the power of
four, i.e. $T\propto a^4$. The binary will hence spend most of its lifetime in
the early phases of the inspiral, to then ``accelerate'' towards the merger and
ringdown. Moreover, the number of cycles that a binary spends is roughly
proportional to the mass ratio of the two compact objects \citep[see,
e.g.][]{Maggiore022018,Maggiore2008}. In the case of an EMRI, this number can
be as large as $10^7$ if we consider a MBH of mass $10^7\,M_{\odot}$ and a
neutron star. This means that an EMRI can spend, as we will see, up to $10^5$
years in the band of the detector if we consider a stellar-mass black hole.

Depending on their evolution stage, i.e. how far they are from plunging through
the event horizon, they will evolve little or nothing in frequency. They are
early-stage EMRIs (E-EMRIs), which have a complete different behavour as
compared to late-type ones, the ones we have been talking about all along.
Depending on their evolution stage, we call them (1) ``monochromatic'', meaning
that the peak of frequencies will not evolve for the observational time (but
obviously the sources will have a spread of harmonics), (2) oligochromatic,
i.e. they explore a short range of frequencies which however is significantly
smaller than that of (3) polychromatic EMRIs, the sources which have been
studied in detail until now in the literature. 

Even though the \textit{merger} event rate (i.e. how many of them cross the
event horizon of the supermassive black hole) can be as low as
$10^{-6}\text{yr}^{-1}$ for our Milky Way \citep{Amaro-Seoane2019}, because of
the time spent in the detection band, what matters is the total number of
E-EMRIs. This can be derived by multiplying the event rate times the lifetime
of the binary with a signal-to-noise ratio (SNR) above a detection threshold,
which we set to 10 in this work.

We find that in our Milky Way, at any given moment, there are hundreds of E-EMRIs
as far as $10^5$ years from plunge fulfilling this requirement, and that they
can achieve SNRs $\gtrsim 10^3$ tens of thousands of years before the plunge.
Using ``realistic'' waveforms (i.e. not just the quadrupole, but based in the
work of \citep{BarackCutler2004a,BarackCutler2004b}), we do a Fisher matrix
parameter study and find that we can obtain the spin and mass of SgrA* with a
negligible errors, as small as $10^{-11}$ and $10^{-5}\,M_{\odot}$,
respectively.  As a side note, we calculate the asymptotic behaviour of the
quadrupole approximation when the eccentricity tends to 1 and find that the
equations in the work of \citep{Peters64} coincides with the asymptotic
solution for eccentricities as large as $e=0.99999$ and higher.

We additionally calculate the incoherent contribution of individual E-EMRIs per
bin of frequency with the proviso that they are continuous. This means that,
from the population of thousands of E-EMRIs, we only select those which have a
continuous signal.  To illustrate this, we consider a population of only two
masses, $10$ and $40\,M_{\odot}$. Because we are taking into account only
continuous signals, from the thousands of sources we are left with about 160 of
them.  These nonetheless are sufficient to build up a foreground signal which
can cover a significant fraction of LISA's sensitivity area, between $10^{-4}$
to $3\times 10^{-3}$ Hz and from $10^{-16}$ to $10^{-21}$ in the characteristic
strain. 

Depending on (1) the mission duration, which affects the binning in frequency,
and (2) the stage in their evolution, some of the sources will be resolved
individually. In this work, however, we are limited to a binning of $10^{-5}$
Hz to depict the incoherent contribution. A narrower binning interval would
render the computational calculations much more intensive and is hence not
included in this study.

It is important to note that discontinuous sources will however also contribute
at the Galactic Centre. These sources will spend only a fraction of their orbit
in the LISA band, and might enter and leave the band in a repeated fashion.
Because of this, we dub them as ``popcorn EMRIs''. Even if they correspond to
non-continuous signals, it is possible to calculate their Fourier transform via
Schwarz functions. We will address these sources and their contribution
elsewhere, in a separate work.

\section{EMRI waveforms with extreme eccentricities: SNR calculation and parameter extraction}

To understand how E-EMRIs located at the Galactic Centre appear in the observatory
band, in this section we first make an analysis by decomposing the approximate
quadrupole waveform into harmonics. The advantage of this technique is that it
gives us relevant information regarding the evolution as a function of
frequency that is hidden if we represent the source with the full waveform.  We
hence approximate the source as a Keplerian ellipse which only changes over
time due to the emission of energy \citep{Peters64}. I.e. it shrinks but we do
not take into account the periapsis shift, nor the effects of the spin of
SgrA*.  However simplified, this first study allows us to understand the
relevance of E-EMRIs. Later we address more complex waveforms which require a
computational analysis.  Thanks to them we perform a Fisher matrix study and
extract the parameters for a few representative examples.

\subsection{A first study via Keplerian ellipses}

From the work of \citep{Peters64} we can do a first analysis if we take into
account that at a given distance $D$, a source of gravitational radiation
emitting a given power $P$ and with a varying frequency $\dot{f}$ has a
characteristic strain which can be expressed as a sum of harmonics $h_{\rm
c,\,n}= {(2\dot E_n/\dot f_n)}^{1/2}/(\pi D)$, as described by
\cite{FinnThorne2000,BarackCutler2004a,BarackCutler2004b}, with $\dot E_n$ the
power radiated to infinity in gravitational waves at a given frequency $f_n =
n\,\nu$, and $n$ an integer larger than 0. Therefore, the contribution of each
harmonic $n$ is given by

\begin{align}
    h_{c,\,n} &= \frac{G^2\,M_{\rm BH} m_{*}}{D\,ac^4} \left(\frac{n^4}{32}\right) \Bigg\{ \Big[J_{n-2}(n e)-2 e J_{n-1}(n e) \nonumber\\
            & +\frac{2}{n} J_{n}(n e)+2 e J_{n+1}(n e)-J_{n+2}(n e)\Big]^{2} \nonumber \\
            & +\left(1-e^{2}\right) \Big[J_{n-2}(n e)-2 J_{n}(n e)+J_{n+2}(n e) \Big]^{2} \nonumber \\
            & +\frac{4}{3 n^{2}}\Big[J_{n}(n e)\Big]^{2} \Bigg\},
\label{eq.hcn}            
\end{align}

\noindent 
with $J_n(x)$ the Bessel functions of the first kind for an argument $x$.

\begin{figure}
\resizebox{\hsize}{!}
          {\includegraphics[scale=1,clip]{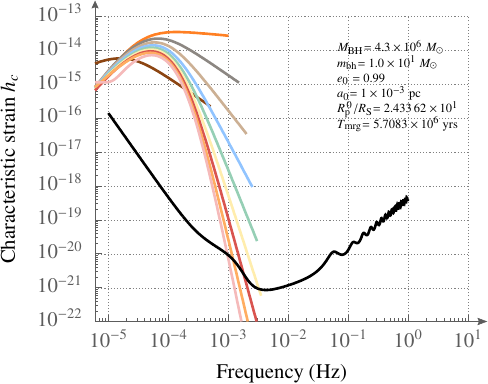}}
\caption
   {
First harmonics of the characteristic amplitude as a function of the frequency
for a standard EMRI \cite[see e.g.][]{Amaro-SeoaneLRR} in the approximation of
\cite{Peters64}. We adopt values for the Galactic Centre, so that
$M_\text{BH}=4.3\times 10^6M_{\odot}$ and the distance $D=8\,\text{kpc}$. We
consider a one-year duration for the mission. The V-shaped, solid line
corresponds to the noise curve $\sqrt{f\,S_h(f)}$ for LISA. On the top, right
corner we show the initial parameters of the binary which we have considered,
which is composed by the massive black hole in our Galaxy, SgrA*, a compact
object of mass $m_*=10\,M_{\odot}$, with initial semi-major axis
$a=10^{-3}\,\text{pc}$ and eccentricity $e=0.997$. Initially, the pericentre of
the orbit is located in this approximation at $R_\text{p}=7.3\,R_\text{S}$,
with $R_\text{S}$ the Schwarzschild radius of SgrA*. At the beginning of the
evolution, the time for the binary to merge due to the emission of
gravitational waves is of $T_\text{mrg}=7.68\times 10^4\,\text{yrs}$. 
    } 
\label{fig.harmonics} 
\end{figure}

In Fig.~(\ref{fig.harmonics}) we depict the first ten harmonics in this
approximation for an EMRI which covers all of the evolutionary phases at the
Galactic Centre, assumed to be at a distance of $D=8$ kpc. The binary enters
the LISA band as a ``monochromatic'' source (and we insist that this is just
terminology, to mean that the peak of frequency will not evolve during the
observational time), to then slowly evolve on timescales of thousands of years
towards the moment in which it circularises, which is when the second harmonic
becomes the dominant one. The E-EMRI will gradually explore more frequencies in
shorter timescales, which is when it becomes oligochromatic, to then cover a
larger range of frequencies when the time to cross the event horizon of the MBH
is of the order of an observational time.  This is the defining characteristic
of polychromatic EMRIs, which makes them interesting to do a ``geo''desic
mapping of warped spacetime. To the best of our knowledge, all EMRIs which have
been discussed in the literature refer only to this last state in the frequency
evolution of an EMRI, when they are polychromatic.

\subsection{Waveforms with extreme eccentricities}

Most used EMRI waveforms are based on the work of \cite{Peters64} and
\cite{PM63}, which adopt Keplerian ellipses shriking over time due to the
emission of gravitational radiation, as in Eq.~(\ref{eq.hcn}).  Usually it is
assumed that a limitation of this derivation is that the eccentricity cannot
be arbitrarily large, since the orbit would be like a pulse and non-integrable.

However, EMRIs naturally form via two-body relaxation which leads to extreme
eccentricities \citep[see e.g.][]{AmaroSeoane2022,Amaro-SeoaneLRR2012}. It is
claimed that none of the existing waveforms can deal with eccentricities
typically larger than $e \gtrsim 0.8$, while in the case of EMRIs we are
looking at $e \gtrsim 0.99$, in particular if the central supermassive black
hole is spinning \citep[][]{Amaro-SeoaneSopuertaFreitag2013}, in which case
they can achieve higher values.  Since a correct calculation of the SNR is
important for the problem that we are addressing, in this section we present a
way to investigate this apparent problem of extreme eccentricities in an
analytical way.

We wish to investigate the behaviour of the solutions to the equations 

\begin{align}
    & \frac{de}{dt} = -\frac{304e}{15}
    \frac{G^3 m_1m_2(m_1+m_2)}{c^5 a^4(1-e^2)^{5/2}}
    \left(1 + \frac{121}{304}e^2\right),\\
    & \frac{da}{de} = \frac{12}{19}\frac{a}{e}
    \frac{\big[1 + (73/24)e^2 + (37/96)e^4\big]}{(1-e^2)\big[1+(121/304)e^2\big]},
\label{eq.AandEPeters}
\end{align}

\noindent 
with the initial condition $e(0)\rightarrow 1$, with $e(0)$ the initial
eccentricity. For $t$ sufficiently small such that $e(t)$ is close enough to 1,
the asymptotic behaviour of these equations is 

\begin{align}
    \label{eqn:de-dt}
    & \frac{de}{dt} = -\frac{85}{3}\frac{G^3m_1m_2(m_1+m_2)}{c^5}\frac{1}{a^4(1-e^2)^{5/2}},\\
    & \frac{da}{de} = \frac{2a}{1-e^2}.
\end{align}

\noindent 
We start by integrating the second equation.

\begin{align}
    \ln a = \ln\left(\frac{1+e}{1-e}\right) + C,
\end{align}

\noindent 
and therefore 

\begin{equation}
    a = C\left(\frac{1+e}{1-e}\right).
\end{equation}

\noindent 
The integration constant $C$ is fixed by the equation

\begin{equation}
    a(0) = C\left(\frac{1+e(0)}{1-e(0)}\right),
\end{equation}

\noindent 
with $a(0)$ the initial semi-major axis.
Setting $a(0)=a_0$ and $e(0)=e_0$ gives us 

\begin{equation}
    \label{eqn:a-e-rel}
    a = \underbrace{a_0\left(\frac{1-e_0}{1+e_0}\right)}_{=C}\left(\frac{1+e}{1-e}\right).
\end{equation}

\noindent 
We now have to calculate the eccentricity as a function of time. Substituting
eq. Eq.~(\ref{eqn:a-e-rel}) to Eq.~(\ref{eqn:de-dt}) gives us 

\begin{align}
    \label{eqn:de-dt-2}
    \frac{de}{dt} = -\frac{1}{\tau} (1-e)^{3/2}(1+e)^{-13/2},
\end{align}

\noindent 
where

\begin{align}
    \frac{1}{\tau} = \frac{85}{3}\frac{G^3m_1m_2(m_1+m_2)}{a_0^4c^5}\left(\frac{1+e_0}{1-e_0}\right)^4.
\end{align}

\noindent 
Again, the equation Eq.~(\ref{eqn:de-dt-2}) is solved by integrating:

\begin{equation}
    \int_{e_0}^e de'\, (1-e')^{-3/2}(1+e')^{13/2} = -\frac{t}{t_0}.
\end{equation}

\noindent 
The indefinite integral on the left term of the equation above is given by 

\begin{align}
    \int de\, & (1-e)^{-3/2} (1+e)^{13/2} = 
    \frac{3003}{8} \tan ^{-1}\left(\sqrt{\frac{1-e}{1+e}}\right) 
    \nonumber\\
    & -\frac{\sqrt{e+1}}{240 \sqrt{1-e}}\left(40 e^6+344 e^5+1406 e^4 \right. \nonumber \\
    & \left. + 3842 e^3+8933 e^2+25499 e-70784\right).
\end{align}

\noindent 
Thus, in the limit $e\rightarrow 1$ the asymptotic behaviour of the integral is 

\begin{align}
   & \int_{e_0}^e de'\, (1-e')^{-3/2}(1+e')^{13/2} = \nonumber \\
   & 128\sqrt{2}\left(\frac{1}{\sqrt{1-e_0}} - \frac{1}{\sqrt{1-e}}\right),
\end{align}

\noindent 
giving us the function of the eccentricity as a function of time:

\begin{equation}
    e(t) = 1 - \frac{1-e_0}{\left(1 + t/t_0\right)^2},
\end{equation}

\noindent 
where 

\begin{align}
    t_0 & = \frac{128\sqrt{2}}{\sqrt{1-e_0}}\,\tau = \nonumber \\
        & \frac{384\sqrt{2}}{85}\frac{G^3 m_1m_2(m_1+m_2)}{c^5}
    \frac{(1-e_0)^{7/2}}{a_0^4 (e_0+1)^4}
\end{align}

The semi-major axis is given by the relation Eq.~(\ref{eqn:a-e-rel}). 

\begin{figure*}
\resizebox{\hsize}{!} 
          {\includegraphics[scale=1,clip]{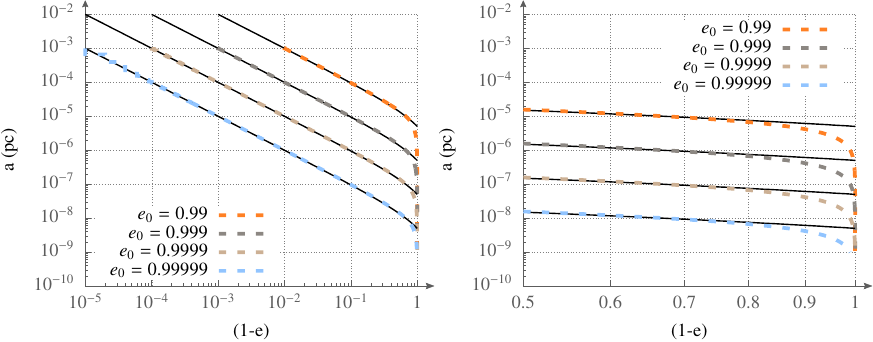}} 
\caption
   {
\textit{Left panel:} Relationship between the semi-major axis and the eccentricity (displayed as $1-e$)
as integrated numerically from Eqs.~(\ref{eq.AandEPeters}), dashed, coloured curves, and the asymptotic
solution presented in this section (solid, black lines). We depict different initial eccentricities. 
   }
\label{fig.Asymptotic_a_e}
\end{figure*}

In Fig.~(\ref{fig.Asymptotic_a_e}) we show this result for the relationship
between the semi-major axis and the eccentricity, and compare it to the
numerical integration of the set of equations given by
Eq.~(\ref{eq.AandEPeters}). As we can see, the asymptotic limit describes the
numerical result well down to eccentricities of about $e \sim 0.2$ This also
confirms that the Peters approximation is a perfectly valid one for
eccentricities of up to $e=0.99999$, since it overlaps with the asymptotic
solution. 

\subsection{Signal-to-noise ratio calculation and parameter extraction}

The SNR of the n-th harmonic can be approximately calculated using

\begin{align}
    \left(\frac{S}{N} \right)^2_n &\approx \frac{h^2_{\rm c,\,n}(f_n)}{5S_h(f_n)}
    \frac{\dot f_n}{f_n^2}T_\mathrm{obs}, \nonumber \\
    \dot{f}_n &\approx n\dot{f}_\mathrm{orb},
\label{eq:snr_approx}
\end{align}

\noindent 
where $T_\mathrm{obs}$ is the observation time and $\dot{f}_\mathrm{orb}$ is
the time derivative of the orbital frequency, which follows from
\cite{Peters64}. To be specific, the square of SNR of the n-th harmonic can be
written as

\begin{align}
    \left(\frac{S}{N} \right)^2_n &\approx \frac{256}{5}\frac{(G\mathcal{M}_c)^{10/3}}{c^8D^2}\frac{g(n,e)}{n^2}(2\pi f_\mathrm{orb})^{4/3}\times \nonumber \\
    & \frac{T_\mathrm{obs}}{5S_h(nf_\mathrm{orb})}.
\label{eq:snr_approx_1}
\end{align}

This way we can see the loudness of each harmonic. This approximation is
inaccurate when the binary is evolving rapidly, when the timescale is shorter
than the observation time.  However, since we are focusing on mono- and
oligochromatic E-EMRIs in this work, this is not a problem. Polychromatic EMRIs
will however need a different treatment.

Therefore, for those cases in which the E-EMRI is close to merger, we generate
the corresponding numerical waveforms and obtain the SNR using

\begin{equation}
    \left(\frac{S}{N}\right)^2=(h|h),
\label{eq:snr_inner_product}
\end{equation}

\noindent
where the inner product of two waveforms $\left(a|b\right)$ is defined in the usual way as
\begin{equation}
	\left(a|b\right):=2\int\mathrm{d}f\ \frac{\tilde{a}(f)\tilde{b}^*(f)+\tilde{a}^*(f)\tilde{b}(f)}{S_h(f)}.
\end{equation}

\begin{figure}
\resizebox{\hsize}{!}
          {\includegraphics[scale=1,clip]{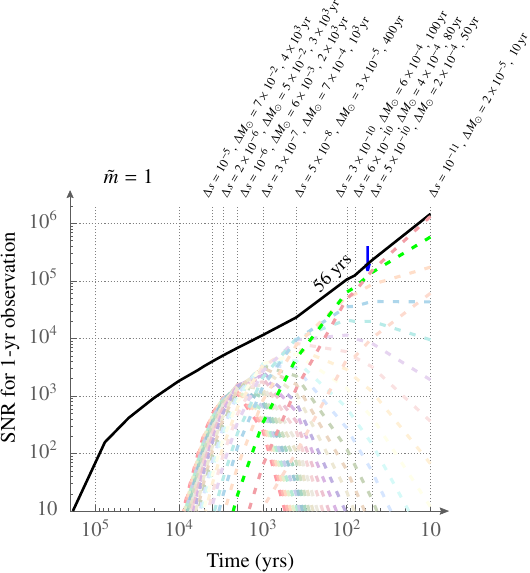}}
\caption
   {
Signal-to-noise calculation for a one-year observation of an E-EMRI at the
Galactic Centre with a mass of $\tilde{m}:=1=10\,M_{\odot}$. The solid, black
line corresponds to the summation of all individual contributions coming from
the different harmonics (dashed, colour curves). The blue arrow shows from
which moment the source is circular, i.e. when the second harmonic dominates
the contribution to the SNR. The vertical, dotted lines show the times at which
the error in the estimation of the spin of the supermassive black hole and its
mass, $\Delta s$ and $\Delta M_{\odot}$, respectively, have been calculated.
Due to the short distance to the source, which is assumed to be located at the
Galactic Centre, the errors are very small, and they become smaller as the
source evolves with time.
   }
\label{fig.SNR_T_10Msun}
\end{figure}

\begin{figure}
\resizebox{\hsize}{!}
          {\includegraphics[scale=1,clip]{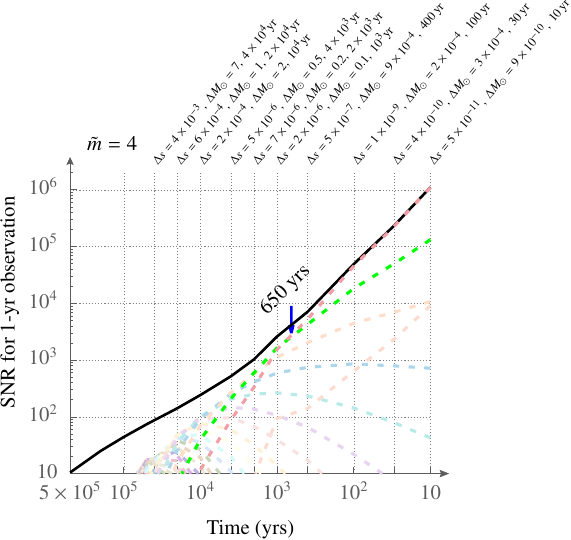}}
\caption
   {
Same as Fig.~(\ref{fig.SNR_T_10Msun}) but for a mass of $\tilde{m}:=4=40\,M_{\odot}$.
   }
\label{fig.SNR_T_40Msun}
\end{figure}

A full comparison between the SNRs obtained in different approach can be found in Table \ref{tab:snr_compare}.

\begin{table}
    \centering
    \begin{tabular}{|c|c|c|c|c|c|}
        \hline
		$T_\mathrm{mrg}$\,({yr}) & $n$ & $\mathrm{SNR_0}$ & $\mathrm{SNR_{aprx}}$ & $\mathrm{SNR'_{aprx}}$ & $\mathrm{SNR_{LW}}$ \\  
		10 & $1-10$ & $1.45\times10^6$ & $4.99\times10^5$ & $4.99\times10^5$ & $3.66\times10^5$ \\
		50 & $1-10$ & $2.27\times10^5$ & $1.24\times10^5$ & $1.24\times10^5$ & $1.25\times10^5$ \\
		80 & $1-10$ & $1.24\times10^5$ & $7.75\times10^4$ & $7.75\times10^4$ & $8.55\times10^4$ \\
		100 & $1-10$ & $1.05\times10^5$ & $6.85\times10^4$ & $6.85\times10^4$ & $7.65\times10^4$ \\
		400 & $1-30$ & $2.71\times10^4$ & $2.32\times10^4$ & $2.32\times10^4$ & $2.89\times10^4$ \\
		2000 & $9-38$ & $7.16\times10^3$ & $6.93\times10^3$ & $7.10\times10^3$ & $9.57\times10^3$ \\
		3000 & $19-48$ & $4.63\times10^3$ & $4.62\times10^3$ & $5.17\times10^3$ & $7.04\times10^3$\\
		4000 & $31-60$ & $3.06\times10^3$ & $3.08\times10^3$ & $3.88\times10^3$ & $5.36\times10^3$\\
        \hline
    \end{tabular}
    \caption{
The SNRs calculated in different approaches. The first column shows the merger
time of the system in years and the second the harmonic number. The 3rd column
'$\mathrm{SNR_0}$' is calculated by generating the waveforms and taking the
square-root of the inner product of the waveform and itself. The Waveforms are
generated with the 10 or 30 loudest harmonics, which are given in the 2nd
column 'n'. The 4th column '$\mathrm{SNR_{aprx}}$' is calculated using the
approximation given in \ref{eq:snr_approx} and considering only the 10 or 30
loudest harmonics given in the 2nd column. The 5th column
'$\mathrm{SNR'_{aprx}}$' uses the same approximation but consider the first
10000 harmonics. The 6th column '$\mathrm{SNR_{LW}}$' is calculated using the
python package LEGWORK\citep{wagg22}.
}
    \label{tab:snr_compare}
\end{table}

Fig.~(\ref{fig.SNR_T_10Msun}) shows the contribution of the first 30 harmonics
of an E-EMRI with a mass of $10\ M_\odot$, and Fig.~(\ref{fig.SNR_T_40Msun}) for
an E-EMRI of mass $40\ M_\odot$.  The SNRs are calculated using Equation
(\ref{eq:snr_approx}), while the SNRs of the cases $\le 100$ yr are calculated
using Equation (\ref{eq:snr_inner_product}).

\begin{table}
    \centering
    \begin{tabular}{|c|c|c|c|c|}
        \hline
		 & $T_\mathrm{mrg}(\mathrm{yr})$ & $R_\mathrm{p}(R_\mathrm{S})$ & $e$ & $a(\mathrm{pc})$ \\ 
		in-band & $1.85\times10^5$ & 7.31 & 0.999333 & $4.52\times10^{-3}$\\
		threshold & $56$ & 4.08 & 0.2244 & $2.17\times10^{-6}$ 
		\\ \hline
    \end{tabular}
    \caption{
Time before merger ($T_\mathrm{mrg}$) in years, pericentre distance in Schwarzschild
radii units ($R_\mathrm{p}$), eccentricity and semi-major axis in
pc of an E-EMRI with a mass of $10\ M_\odot$ at two specific moments: (1)
in-band, the moment when the SNR of this system reaches 10; (2) threshold, the
moment from which the source becomes circular, i.e.  the second harmonic
dominates. These data are used later for the calculation of number of sources
in band when they are eccentric and circular.
}
    \label{tab:inband_and_threshold_10}
\end{table}

\begin{table}
    \centering
    \begin{tabular}{|c|c|c|c|c|}
        \hline
		 & $T_\mathrm{mrg}(\mathrm{yr})$ & $R_\mathrm{p}(R_\mathrm{S})$ & $e$ & $a(\mathrm{pc})$ \\ 
		in-band & $5\times10^5$ & 23.93 & 0.959765 & $2.45\times10^{-4}$\\
		threshold & $650$ & 11.47 & 0.1541 & $5.59\times10^{-6}$ 
		\\ \hline
    \end{tabular}
    \caption{
Same as Table~(\ref{tab:inband_and_threshold_10}) but for a mass of $40\,M_\odot$.
}
    \label{tab:inband_and_threshold_40}
\end{table}

\begin{figure}
\resizebox{\hsize}{!} 
          {\includegraphics[scale=1,clip]{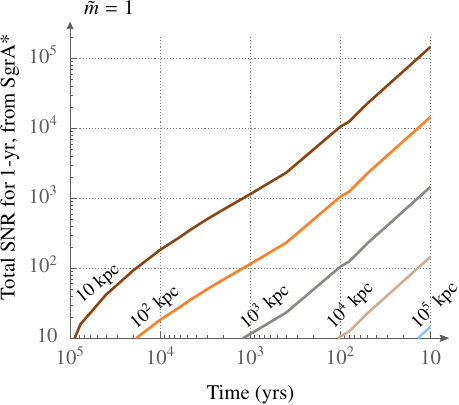}} 
\caption
   {
SNR for sources located at different distances from our Galaxy, assuming a 
one-year observation time. We can see that LISA can in principle observe E-EMRIs 
(of mass $10\,M_{\odot}$) in galaxies out to distance of approximately 0.1 Gpc.
   }
\label{fig.SNR_T_10Msun_DifferentD}
\end{figure}

We perform a Fisher matrix analysis to see how precise we can measure the mass
and the spin of the SMBH with a single event. 

The Fisher matrix can be defined as
\begin{equation}
    \Gamma_{ij}=\left(\frac{\partial h(\mathbf{\theta})}{\partial\theta_i}\Big|\frac{\partial h(\mathbf{\theta})}{\partial\theta_j}\right),    
\end{equation}
where $h(\mathbf{\theta})$ is the waveform of an EMRI and $\mathbf{\theta}$ is a vector containing the parameters (e.g. the mass and the spin of the SMBH).
To obtain the partial derivative numerically, we calculate two waveforms whose parameter vectors $\mathbf{\theta}$ only differ from each other in the $i$-th element: $(\theta_i+\delta_i/2)$ and $(\theta_i-\delta_i/2)$. 
When $\delta_i$ is adequately small, the partial derivative is approximately
\begin{equation}
    \frac{\partial h(\mathbf{\theta})}{\partial\theta_i}
    =\frac{h(\theta_i+\delta_i/2)-h(\theta_i-\delta_i/2)}{\delta_i}.
\end{equation}
The accuracy (or error) of the $i$-th parameter will be the square root of the $i$-th diagonal element of the inverse of the Fisher matrix 
\begin{equation}
    \Delta\theta_i=\sqrt{\left(\Gamma^{-1}\right)_{ii}}
\end{equation}

The results at different time are
listed in Table~(\ref{tab:snr_and_para_10}) for the $10\,M_{\odot}$ source and
Table (\ref{tab:snr_and_para_40}) for the $40\,M_{\odot}$ one.  The numerical analysis is performed to generate the requisite waveforms.    
Their generation can
be time-consuming when the eccentricity of the orbit is very large, since more
harmonics contribute and must hence be calculated.  Due to limited computation
time, we have only calculated those cases close to merger (with low
eccentricity).  The case with the lowest orbital frequency in both tables is
the last line in Table \ref{tab:snr_and_para_40} (corresponding to $40\
M_\odot$ and $4\times10^4$ years before plunge), in which the compact object
can still make $>100$ orbits in 1-yr observation time and thus guarantees the
reliability of the Fisher matrix analysis.

\begin{table}
    \centering
    \begin{tabular}{|c|c|c|c|c|c|}
        \hline
		$T_\mathrm{mrg}$\,($\mathrm{yr}$) & $e$ & $R_\mathrm{p}^{\,0}(R_\mathrm{S})$ & $n$ & $\Delta$ s & $\Delta M$\,($M_\odot$) \\ 
		10 & 0.1216 & 3.0 & $1-10$ & $1\times10^{-11}$ & $2\times10^{-5}$\\
		50 & 0.21515 & 4.0 & $1-10$ & $5\times10^{-10}$ & $2\times10^{-4}$\\
		80 & 0.25489 & 4.4 & $1-10$ & $6\times10^{-10}$ & $4\times10^{-4}$\\
		100 & 0.27592 & 4.5 & $1-10$ & $3\times10^{-10}$ & $6\times10^{-4}$\\
		400 & 0.45551 & 5.5 & $1-30$ & $5\times10^{-8}$ & $3\times10^{-5}$\\
		1000 & 0.60832 & 6.2 & $3-32$ & $3\times10^{-7}$ & $7\times10^{-4}$\\
		2000 & 0.72767 & 6.5 & $9-38$ & $1\times10^{-6}$ & $6\times10^{-3}$\\
		3000 & 0.79255 & 6.7 & $19-48$ & $2\times10^{-6}$ & $5\times10^{-2}$\\
		4000 & 0.83423 & 6.9 & $31-60$ & $1\times10^{-5}$ & $7\times10^{-2}$\\
        \hline
    \end{tabular}
    \caption{Parameter extraction via Fisher matrix analysis of an E-EMRI with a mass of 
             $10\ M_\odot$. From the left to the right we show the time for merger in
             years, the eccentricity, the initial periapsis distance, the loudest harmonics
             used in the generation of the waveforms, the error in the spin measurement and
             the error in the measurement of the mass of the MBH, SgrA*.}
    \label{tab:snr_and_para_10}
\end{table}

\begin{table}
    \centering
    \begin{tabular}{|c|c|c|c|c|c|}
        \hline
		$T_\mathrm{mrg}$\,($\mathrm{yr}$) & $e$ & $R_\mathrm{p}^{\,0}(R_\mathrm{S})$ & $n$ & $\Delta$ s & $\Delta M$\,($M_\odot$) \\ 
		10 & 0.029 & 4.5 & $1-10$ & $5\times10^{-11}$ & $9\times10^{-10}$\\
		30 & 0.046 & 5.9 & $1-10$ & $4\times10^{-10}$ & $3\times10^{-4}$\\
		100 & 0.074 & 7.7 & $1-10$ & $1\times10^{-9}$ & $2\times10^{-4}$\\
		400 & 0.130 & 10.5 & $1-10$ & $5\times10^{-7}$ & $9\times10^{-4}$\\
		1000 & 0.180 & 12.4 & $1-10$ & $2\times10^{-6}$ & $1\times10^{-1}$\\
		2000 & 0.237 & 14.1 & $1-15$ & $7\times10^{-6}$ & $2\times10^{-1}$\\
		4000 & 0.311 & 15.9 & $1-15$ & $5\times10^{-6}$ & $5\times10^{-1}$\\
		$1\times10^4$ & 0.430 & 18.1 & $1-20$ & $2\times10^{-4}$ & $2$\\
		$2\times10^4$ & 0.538 & 19.7 & $1-30$ & $6\times10^{-4}$ & $1$\\
		$4\times10^4$ & 0.692 & 21.4 & $8-37$ & $4\times10^{-3}$ & $7$\\
        \hline
    \end{tabular}
    \caption{Same as Table~(\ref{tab:snr_and_para_10}) but for a mass of $40\ M_\odot$}
    \label{tab:snr_and_para_40}
\end{table}

\section{Number of E-EMRIs in band at the Galactic Centre}

In order to calculate the number of these sources in band, we adopt the result
of \cite{Amaro-Seoane2019}, which can be directly applied to our problem. In
his work, the author derives the event rate ${\Gamma}$ for merging stellar-mass
black holes with supermassive black holes as a test in his Eq.~(40),

\begin{align}
{\Gamma} & \sim 1.92\times 10^{-6}\,\textrm{yrs}^{-1}\tilde{N}_{0}\,\tilde{\Lambda}\,\tilde{R}_{0}^{-2}\,\tilde{m}^2 \times \nonumber\\
                           & \Bigg\{
                                    1.6\times 10^{-1} \tilde{R}_{0}^{1/2}\tilde{N}_{0}^{-1/2}\tilde{\Lambda}^{-1/2}\tilde{m}^{1/2}\,{\cal W}(\iota,\,{\rm s})^{-5/4}\times \nonumber\\
                           &         \left[\ln\left(9138\, \tilde{R}_{0}\, \tilde{N}_{0}^{-1} \tilde{\Lambda}^{-1}\tilde{m}\,{\cal W}(\iota,\,{\rm s})^{-5/2}\right) - 2 \right] - \nonumber \\
                           &         4\times 10^{-2}\tilde{R}_{0}^{1/2}\times \left[\ln\left(618\, \tilde{R}_{0}\right) - 2 \right]
                             \Bigg\},
\label{eq.eventrates}
\end{align}

\noindent 
with the following notation,

\begin{align}
\tilde{\Lambda}& :=\left(\frac{\ln(\Lambda)}{13}\right),~ \tilde{N}_{0}:=\left(\frac{N_0}{12000}\right)\nonumber \\
\tilde{R}_{0}& :=\left(\frac{R_{\rm h}}{1\textrm{pc}}\right),~ \tilde{m}:=\left(\frac{m}{10\,M_{\odot}}\right).
\end{align}

\noindent 
Here $\ln(\Lambda) \simeq \ln(M_{\rm BH}/m)$, with $M_{\rm BH}$ the mass of the MBH,
$m$ the mass of the stellar-mass black hole
\cite[see][]{Amaro-SeoaneLRR2012} and $N_0$ the number of stellar-mass black
holes within a radius $R_0$ where relaxation can be enviaged to be completely
dominated by this kind of compact objects.  In the same work,
\cite{Amaro-Seoane2019} explains why $R_0$ should be set to the influence radius $R_{\rm h}$
(which we accordingly use in the numerator of the equation).  The function
$\cal{W}(\iota,\,{\rm s})$ reflects the location of the separatrix of a Kerr
MBH as compared to a Schwarzschild case, and depends on the magnitude of the
spin of the MBH, $\rm{s}$ and the inclination of the orbit, $\iota$ \citep{Amaro-SeoaneSopuertaFreitag2013}.

\begin{figure*}
\resizebox{\hsize}{!} 
          {\includegraphics[scale=1,clip]{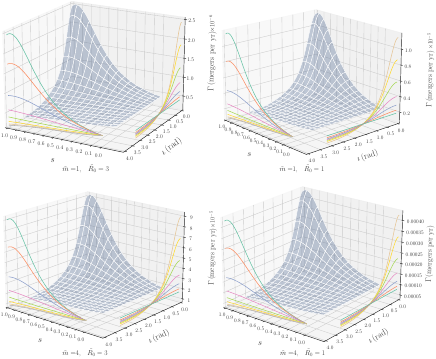}}
\caption
   {
\textit{Upper panels:} Event rates, i.e. mergers, as derived in Eq.~(\ref{eq.eventrates}) for a stellar-mass black
hole of mass $\tilde{m}=1$, which corresponds to $10\,M_{\odot}$ for an influence radius of 3pc (left
panel) and 1 pc (right panel). We depict the rates as a function of the value of the spin of the 
supermassive black hole, whose mass is set to $4.3\times10^6\,M_{\odot}$ and the inclination of the
orbit in radians. 
\textit{Lower panels:} Same but for a mass of $40\,M_{\odot}$.
   }
\label{fig.Gamma}
\end{figure*}

We note that Eq.~(\ref{eq.eventrates}) is self-contained. This means that it
contains the information about the integration limits to derive the event
rates; i.e. the number of stellar-mass black holes plunging through the event
horizon of the MBH per year.  These two limits are the minimum radius, the
distance within which we expect to have at least one object to start the
integration, and the threshold radius which divides the relaxation-dominated
regime and the one in which the driving mechanism is the emission of
gravitational waves \cite[see][for a more detailed explanation and
derivation]{Amaro-Seoane2019,Amaro-SeoaneLRR2012}.  This radius is given in
Eq.(28) of \cite{Amaro-Seoane2019} in a general expression.  In the particular
case of a stellar-mass black hole, this radius is

\begin{align}
a_{\rm crit} &=  R_{0} \Bigg[\frac{20480}{1207} (3-\gamma)(1+\gamma)^{3/2}  \times \nonumber \\
               &                   \times  
                                           C\,{\cal W}(\iota,\,{\rm s})^{5/2}\,N_{0} \ln(\Lambda)
                                           \left(\frac{M_{\rm BH}}{m}\right)^{-1}
                                            \Bigg]^{\frac{1}{\gamma-3}},
\label{eq.acrit}
\end{align}

\noindent
where $C$ is a factor of order 1 and $\gamma$ is the exponential of the power-law cusp
that the stellar-mass black holes form around the MBH. For an initial-mass function 
which represents our Galaxy, this is $\gamma=2$ \citep{PretoAmaroSeoane10,AlexanderHopman09}.

Hence, we can use Eq.~(\ref{eq.eventrates}) for any mass of stellar-mass black
hole that we want to address. We only need to choose the mass of the object,
since $\tilde{\Lambda}$ and $\tilde{N_0}$ are the same for all masses. With the
choice of the mass, the rest of the parameters is accordingly fixed.

In order to obtain the number of sources in band, we follow exactly the same
argument we introduced in \cite{Amaro-Seoane2019}, and we therefore refer the
reader to that work for details. Although in that work we were focusing on a
different kind of source, the physical idea remains the same. We succinctly
summarise the fundamentals.  One could think that the total number of sources
in band can be obtained by simply multiplying the event rate by the time spent
in band. The total number of sources that the observatory is to detect can be
derived by solving a differential equation. We know the influx of new sources
from the stellar system thanks to relaxation theory, which allows us to define
the critical radius which we presented before, and we know how often one of
them disappears from the system through the sink, since we have the rates, i.e.
how often it merges with the central MBH. By multiplying the number of sources
per year by the time spent on band, we should obtain the number of sources at
any given time.

This argument is however not correct, since the energy of the E-EMRIs sources
must be below a threshold value in phase-space, as specified in
\cite{Amaro-Seoane2019}. It is not enough to diffuse in energy to achieve
semi-major axis values below $a_\text{crit}$ as given by Eq.~(\ref{eq.acrit}).
We recall that $a_\text{crit}$ merely marks the transition in the value of the
semi-major axis from the regime in which dynamics dominates the evolution of
the binary to the one dominated by general relativity (see Fig.~5 of
\cite{Amaro-Seoane2019}). From that value, the binary still has a long way to
do until it enters the band of the detector. I.e. the orbital period must
become much shorter. Not all sources crossing the critical radius will
successfully arrive to that point.

Therefore, E-EMRIs approach must first shrink their semi-major axes to values
lower than $a_\text{crit}$, and only a fraction of them will arrive to a
value which we call $a_\text{band}$, in which the source is in band. From there
they will slowly circularise and cross $a_\text{thr}$ to finally cross the minimum
semi-major axis which leads to a merger, $a_\text{min}$.

In physical space, finding which sources do arrive corresponds to solving in
the limit in which $\Delta t \to 0$ the following continuity equation for a
line density function $l \equiv dN/da$ with $N$ the number of sources with a
given energy

\begin{equation}
\frac{\partial}{\partial a} \Big[\, \dot{a}(a,\,e)\,l \,\Big] + \frac{\partial l}{\partial t} =0.
\label{eq.cont}
\end{equation}

Following \cite{Amaro-Seoane2019}, we now introduce two different regimes,
which will allow us to normalise $N$ and obtain the total number of sources. We
thus distinguish the circular and the eccentric regime and a threshold distance
$a_{\rm thr}$ which marks that transition. In the eccentric regime, we will
approximate the eccentricity to be $e \sim 1$ and vice-versa, in the circular
regime we will take $e \sim 0$. In practice, $a_{\rm thr}$ can be determined by
finding out when the second harmonic dominates over the rest of them. We have
this information from the previous section.  We can take a representative case,
for instance that of Fig.~(\ref{fig.harmonics}), in which this transition
happens at a pericentre distance of $R_\text{p} = 3.67\,R_\text{S}$ and an
eccentricity of $e \sim 0.18$. Therefore we have that $a_{\rm
thr}=R_\text{p}/(1-e) \sim 4.5\,R_\text{S} \sim 2\times 10^{-6}\,\text{pc}$,
since for $M_\text{BH}=4.3\times10^6\,M_{\odot}$, we have $R_\text{S} \sim
4\times 10^{-7}\,\text{pc}$. On the other hand, the minimum semi-major axis can
be defined to be $a_\text{min} = 2 \times R_\text{S} \sim 8 \times
10^{-7}\,\text{pc}$.

Let us call the number of sources comprised between $a_\text{min}$ and
$a_\text{thr}$, $N_\text{I}$, the number comprised between $a_\text{thr}$ and
$a_\text{band}$, $N_\text{II}$ and from there to $a_\text{crit}$,
$N_\text{III}$ (as in Fig.~7 of \cite{Amaro-Seoane2019}). From the same work, the
solution to the differential equation is

\begin{align}
\frac{N_{\rm II}}{N_{\rm III}}           & = \frac{a_{\rm band}^{1/2}-a_{\rm thr}^{1/2}}
                                                  {a_{\rm crit}^{1/2}-a_{\rm band}^{1/2}} \nonumber \\
\frac{N_{\rm I}}{N_{\rm II}+N_{\rm III}} & = \frac{1}{8} \times 
                                             \frac{1-\left(a_{\rm min}/a_{\rm thr}\right)^4}
                                             {\left(a_{\rm crit}/a_{\rm thr} \right)^{1/2}-1}\nonumber \\
N_{\rm I}+N_{\rm II}                     & = {\Gamma} \times T\left(a,\,e\right),
\end{align}

\noindent 
with $T\left(a,\,e\right)$ the amount of time spent on band with a minimum SNR of 10.

\noindent 
Therefore, the full expression for each number of E-EMRIs per interval is given
by the expressions

\begin{align}
N_{\rm I}   & = \frac{CD(1+A) }{A + (1+A)D}\nonumber \\
N_{\rm II}  & = \frac{CA}{A + (1+A)D} \nonumber \\
N_{\rm III} & = \frac{C}{A + (1+A)D},
\end{align}

\noindent 
with

\begin{align}
A & \equiv \frac{a_{\rm band}^{1/2}-a_{\rm thr}^{1/2}}{a_{\rm crit}^{1/2}-a_{\rm band}^{1/2}} \nonumber \\
D & \equiv \frac{1}{8} \frac{1-\left(a_{\rm min}/a_{\rm thr}\right)^4} {\left(a_{\rm crit}/a_{\rm thr} \right)^{1/2}-1}\nonumber \\
C & \equiv {\Gamma} \times T\left(a,\,e\right).
\end{align}

\noindent 
More explicitly,

\begin{align}
N_{\rm I}   & = {\Gamma} \times T\left(a,\,e\right) \times \Omega_1 \nonumber \\
N_{\rm II}  & = {\Gamma} \times T\left(a,\,e\right) \times \Omega_2 \nonumber \\
N_{\rm III} & = {\Gamma} \times T\left(a,\,e\right) \times \Omega_3,
\end{align}

\noindent 
where we have introduced the weighting functions $\Omega_1$, $\Omega_2$ and $\Omega_3$ given in
Eqs.~(\ref{eq.Omegas})

\begin{figure*}
\begin{align}
\Omega_1 & \equiv \frac{(\sqrt{a_{\rm thr}} - \sqrt{a_{\rm crit}})(a_{\rm min}^4 - a_{\rm thr}^4)}{(a_{\rm thr}^4 (8 \sqrt{a} (\sqrt{a_{\rm crit}/a_{\rm thr}} - 1) + \sqrt{a_{\rm crit}}) + a_{\rm thr}^{9/2} (7 - 8 \sqrt{a_{\rm crit}/a_{\rm thr}}) + a_{\rm min}^4 (\sqrt{a_{\rm thr}} - \sqrt{a_{\rm crit}})} \nonumber \\
\Omega_2 & \equiv \frac{8 a_{\rm thr}^4 (\sqrt{a} - \sqrt{a_{\rm thr}}) (\sqrt{a_{\rm crit}/a_{\rm thr}} - 1)}{a_{\rm thr}^4 (8 \sqrt{a} (\sqrt{a_{\rm crit}/a_{\rm thr}} - 1) + \sqrt{a_{\rm crit}}) + a_{\rm thr}^{9/2} (7 - 8 \sqrt{a_{\rm crit}/a_{\rm thr}}) + a_{\rm min}^4 (\sqrt{a_{\rm thr}} - \sqrt{a_{\rm crit}})} \nonumber \\
\Omega_3 & \equiv \frac{8 a_{\rm thr}^4 (\sqrt{a} - \sqrt{a_{\rm crit}}) (1 - \sqrt{a_{\rm crit}/a_{\rm thr}})}{(a_{\rm thr}^4 (8 \sqrt{a} (\sqrt{a_{\rm crit}/a_{\rm thr}} - 1) + \sqrt{a_{\rm crit}}) + a_{\rm thr}^{9/2} (7 - 8 \sqrt{a_{\rm crit}/a_{\rm thr}}) + a_{\rm min}^4 (\sqrt{a_{\rm thr}} - \sqrt{a_{\rm crit}})}
\label{eq.Omegas}
\end{align}
\end{figure*}

\begin{figure*}
\resizebox{\hsize}{!} 
          {\includegraphics[scale=1,clip]{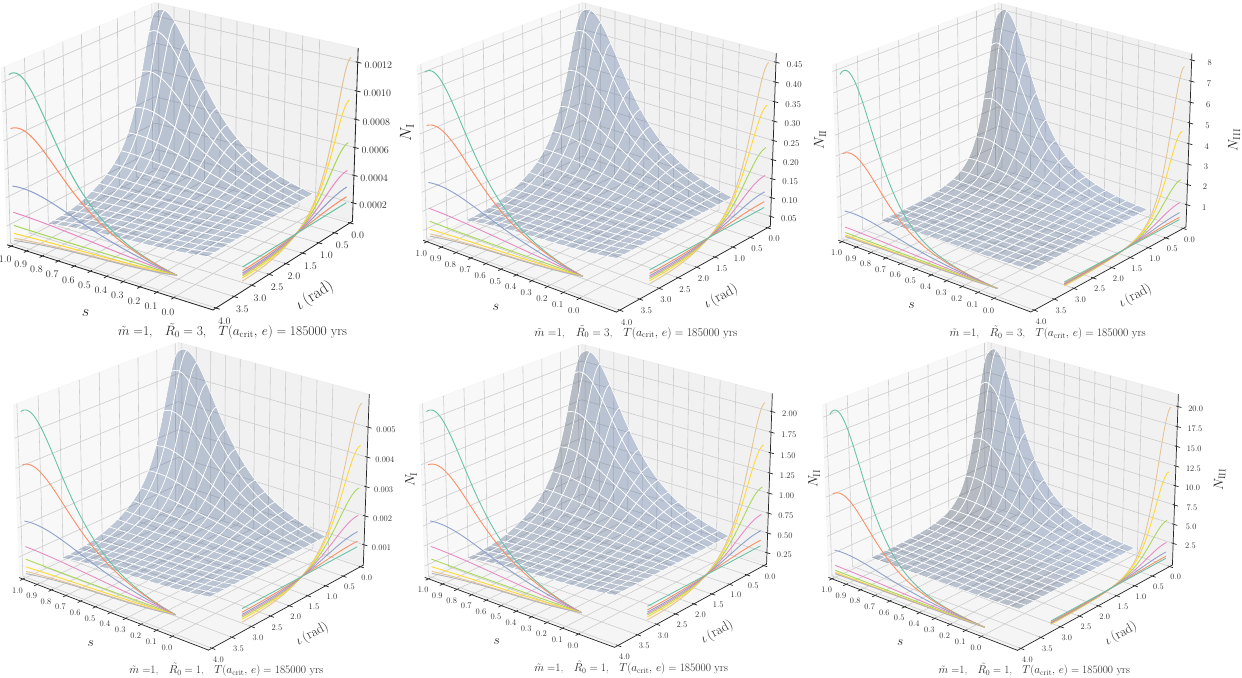}}
\caption
   {
Number of E-EMRIs $N({\rm s},\,\iota)$ as a function of the
inclination of the orbit $\iota$ and the spin of the MBH, s, for a stellar-mass
black hole of mass $10\,M_{\odot}$ and two influence radii, 3pc (upper panels)
and 1pc (lower panels). From the left to the right we have the number of
sources for the regions $N_{\rm I}$, $N_{\rm II}$ and $N_{\rm III}$.
   }
\label{fig.N_10Msun}
\end{figure*}

In Fig.~(\ref{fig.N_10Msun}) we display the result for a stellar-mass black
hole of mass $10\,M_{\odot}$ and different influence radii. The value of this
parameter has an important impact in the total number of E-EMRIs in the three
different regions, as we can see. Observations of the Galactic Centre derive a
value of $R_\text{infl}=3\,\text{pc}$ \cite{SchoedelEtAl2018}, but we also add
the results for $R_\text{infl}=1\,\text{pc}$, which has been the ``default''
value in the related literature. The number of E-EMRIs in the oligochromatic
phase, i.e. in the range of frequencies close to that of a polychromatic EMRI,
are negligible, of the order of $10^{-3}$ for both values of the influence
radius. As we move to lower frequencies and get into the monochromatic stage,
we can expect to have two sources if the influence radius is one, with
associated SNR values around $10^5$, as we can see in
Fig.~(\ref{fig.SNR_T_10Msun}). In the farthest region, where the sources are
more eccentric, the number of E-EMRIs lies between 8 and 20, with SNRs that are
smaller in comparison but still around thousands.

\begin{figure*}
\resizebox{\hsize}{!} 
          {\includegraphics[scale=1,clip]{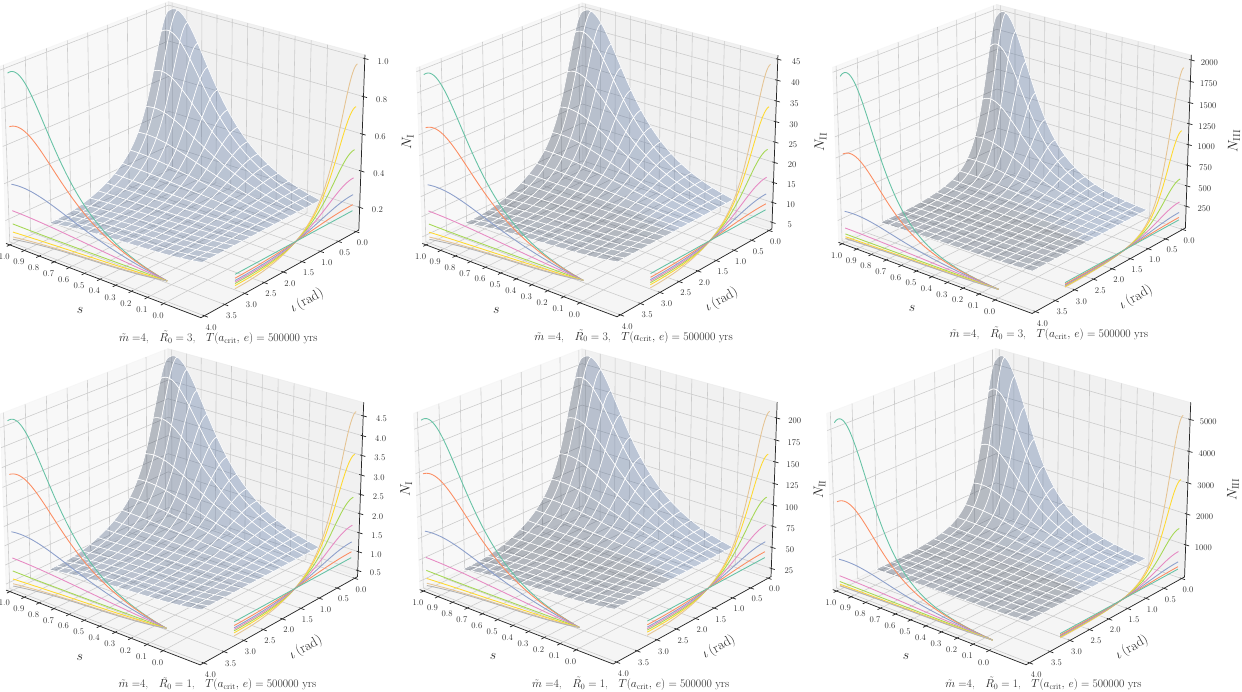}}
\caption
   {
Same as Fig.~(\ref{fig.N_10Msun}) but for a stellar-mass black hole of mass
$40\,M_{\odot}$.
   }
\label{fig.N_40Msun}
\end{figure*}

We also show the results for a different mass value of the stellar-mass black
hole, $40\,M_{\odot}$, and different influence radii, in
Fig.~(\ref{fig.N_40Msun}). In this case, and as we expected from the larger
event rates and time spent on band, the numbers are more interesting. In region
$N_\text{I}$, where we can expect E-EMRIs to get SNR values close to $10^6$, as
we can see in Fig.~(\ref{fig.SNR_T_40Msun}), we expect to have between one or a
handful of sources, depending on the value of $R_\text{infl}=3\,\text{pc}$. In
region $N_\text{II}$, the numbers range between 45 and 200, and the SNR is
typically of the order $10^4$. Finally, in region $N_\text{I}$, where the SNR
is the lowest, but still can reach values as high as thousands, we predict
between 2000 and 5000 sources.

It is important to stress out that these numbers are not to be envisaged as
rates of stellar-mass black holes crossing the event horizon. They are telling
us how many E-EMRIs are at the Galactic Centre as the result of a steady-state
analysis. I.e. they represent on average the number of E-EMRIs which can be
found in galactic nuclei (with the representative parameters we have addressed)
\textit{at any given moment} in the evolution of that nucleus. The proviso is
that the nucleus is relaxed, which is true if the mass of the MBH has at most a
few times $10^7\,M_{\odot}$ \cite{AmaroSeoane2022}, but also that the MBH is fixed
at the centre of the system: for intermediate-mass black holes the derivation would
be significantly more complex, since it would be wandering around the centre.

\section{A forest of E-EMRIs}

In this section we estimate what the foreground signal of \textit{continuous}
E-EMRIs would be. There are two important assumptions we are making. First, the
binning in frequencies is broader than what in reality LISA will be. This is so
because the computational estimation of thinner binnings is too demanding.  The
implication is that we are assuming that the sources are contributing in an
incoherent way. As mentioned in the introduction, this is not necessarily the
case: A narrower binning might allow us to distinguish individual sources, in
particular when they are far away from their polychromatic behaviour, i.e. when
they are ``monochromatic''. The second one is that we consider only continuous
signals, and hence from the thousands mentioned previously we are left with
only about 160 of them for the $10\,M_{\odot}$, as we explain later. In
reality, however, there will be sources contributing to the signal which are
not continuous. However, discontinuous sources can be dealt with, as we mention
later.

\begin{figure}
\resizebox{\hsize}{!}
          {\includegraphics[scale=1,clip]{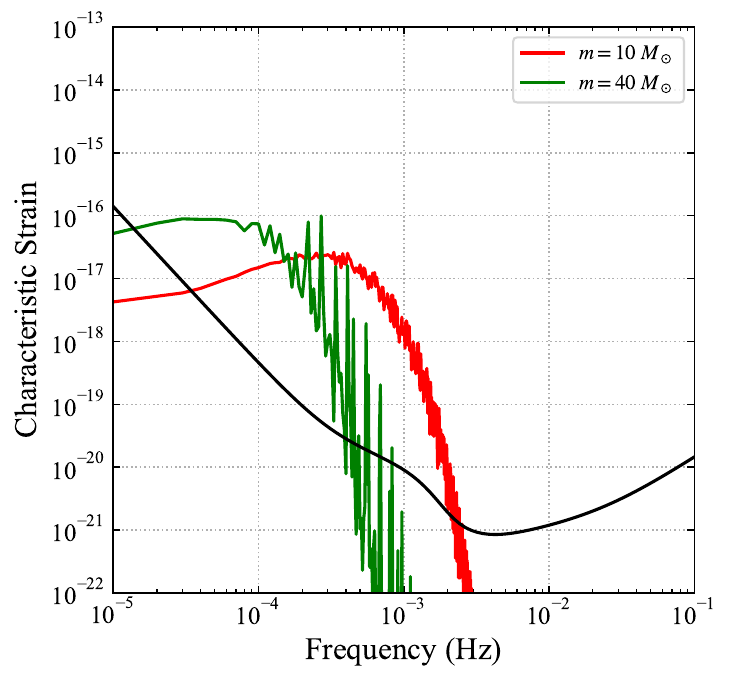}}
\caption
   {
Total contribution of of E-EMRIs at the Galactic Centre during 1-yr observation
time.  The two different colours indicates different masses of the E-EMRIs,
assumed to be of either $10\,M_{\odot}$ or $40\,M_{\odot}$. As we
move towards higher frequencies, the signal becomes less smooth
because there are less and less individual sources. On the other
hand, the closer to merger, the larger the SNR, which explains
why the characteristic amplitude does not drop as quick as one
would naively expect. 
   }
\label{fig.emri1000}
\end{figure}

The characteristic strain of the E-EMRI background $h_{c,\mathrm{gwb}}$ can be calculated using
the following expression

\begin{equation}\label{eq:bgcalc_origin}
    \begin{aligned}
    h^2_{c,\mathrm{gwb}}(f)=\frac{1}{2}&\int\mathrm{d}\mathcal{M}\ \mathrm{d}e
    \left[\sum_n\frac{\mathrm{d}^4N}
    {\mathrm{d}\mathcal{M}\ \mathrm{d}e\ \mathrm{d}\ln{f_\mathrm{orb}}}
    \times\frac{h_{c,n}^2}{fT_\mathrm{obs}}\right],\\
    f_\mathrm{orb}&=\frac{f(1+z)}{n}\in[f_\mathrm{orb,min},f_\mathrm{orb,max}],
    \end{aligned}
\end{equation}

\noindent 
where we have followed the same nomenclature as in \citep{BonettiSesana2020};
i.e.  $\mathcal{M}$ is the the source-frame chirp mass, $f_\mathrm{orb,min}$
and $f_\mathrm{orb,max}$ are the minimum and maximum orbital frequencies and
$z$ the redshift (which in our case is zero, since we are focusing on the
Galactic Centre).

We depict the result in Fig.~(\ref{fig.emri1000}). The orbital parameters
(eccentricity and orbital frequency) of these E-EMRIs are selected as follows: (1) We
pick an E-EMRI that has just entered the LISA band; (2) we then evolve it
until 1 year before plunge (the assumed observational time); (3) we randomly
choose different instants from the above evolution process, which allows us to
interpret them as different E-EMRIs instead of calculating different waveforms
for each source; finally, (4) we calculate their characteristic strain and sum
them up.  We do such calculations both for the mass of the stellar-mass black
hole $m=10~M_\odot$ and $m=40~M_\odot$ (shown as red and green line in Figure
\ref{fig.emri1000}). The mass of the MBH is fixed to
$M_\mathrm{BH}=4.3\times10^6\ M_\odot$.  The initial parameters are
$e=0.983005,~a=1.76376\times10^{-4}~\rm pc$ for $m=10~M_\odot$ and
$e=0.959765,~a=2.45\times10^{-4}~\rm pc$ for $m=40~M_\odot$. 

The initial parameters above correspond to the in-band parameters in Tables
(\ref{tab:inband_and_threshold_10}) and (\ref{tab:inband_and_threshold_40}).
We note that in the $10~M_\odot$ case we adopt the parameters of an E-EMRI
$3.03\times10^4~\rm yr$ before plunge. From the total initial number we select
only continuous E-EMRIs, which decreases the total number to $\sim 163$ in the
$10\,M_{\odot}$. As mentioned before, this is due to limited computation time,
since E-EMRIs at their early age have smaller orbital frequency and large
eccentricity, and therefore require more harmonics to be considered in the
calculation, which causes the computation time to grow rapidly. In the
$40\,M_{\odot}$ case, all of the initial cases were kept, i.e. they were
continuous.

The LISA band is divided into bins of width $\delta f\sim10^{-5}\
\mathrm{Hz}$).  For each E-EMRI in the sample, we calculate the characteristic
strain of all the harmonics in LISA band ($10^{-5}\
\mathrm{Hz}<nf_\mathrm{orb}<10^{-1}\ \mathrm{Hz}$) and add their contribution
to the corresponding bins.  The contribution of the $n$-th harmonic of the
$m$-th E-EMRI $(h^2_{c,\mathrm{gwb}})_{m,n}$ is calculated with

\begin{equation}
    (h^2_{c,\mathrm{gwb}})_{m,n}=
    \left\{
    \begin{aligned}
    &\frac{h^2_{c,n}}{2fT_\mathrm{obs}} \quad 
        &\mathrm{if}\ \Delta f_{m,n}>\delta f \\
    &\frac{f h^2_n}{2\ \delta f}
        &\mathrm{if}\ \Delta f_{m,n}\le\delta f,
    \end{aligned} 
    \right.
\end{equation}

\noindent 
where $f=n f_{\mathrm{orb},m}$ is the frequency of this harmonic and $\Delta
f_{m,n}$ is the change of the frequency during the observation time of
$T_\mathrm{obs}=1~\mathrm{yr}$. If $\Delta f_{m,n}>\delta f$, we would assume
that this harmonic affects more than one frequency bin, and contributes noise
to all the bins related. Otherwise, its contribution is limited to only one
bin.  We repeat the above procedure five times and average the results to
ensure the statistical reliability.

\section{Conclusions}

In this work we have estimated the number of sources which are on their way to
becoming EMRIs at our Galactic Centre.  The number of them crossing the event
horizon, i.e. the mergers, is very low, as expected. However, we show that at
any given time (meaning that this is a steady-state solution, with the proviso
that the relaxation time is shorter than the Hubble time and the MBH is fixed
at the centre) there is a large number of early-stage EMRIs (which we dub
E-EMRIs), i.e. EMRIs which yield a SNR of at least 10 in the LISA observatory
for a one year observation time. The interesting feature is that E-EMRIs can
spend up to $5\times10^5$ years in band with a SNR starting at 10. 

Using waveforms following the scheme of
\citep{BarackCutler2004a,BarackCutler2004b}, we perform a SNR calculation and a
Fisher matrix study in order to extract parameters. We also perform an
asymptotic analysis of the equations when $e \to 1$ and find that the treatment
of the eccentricity for extreme values is correct.

Due to the short distance to the Galactic Centre, we find SNR ranging between a
minimum of 10 to up to $10^6$ depending on their evolutionary stage.  Because
of the large SNRs, E-EMRIs can be detected in other galaxies, out to 0.1 Gpc.
In our parameter study we focus on the spin and mass of the MBH of our Galactic
Centre and find errors as small as $10^{-11}$ and $10^{-5}\,M_{\odot}$
respectively.

We distinguish three different kinds of E-EMRIs depending on their evolutionary
phases so as to derive the number of them based on a normalisation of their
orbital parameters by integrating a line density function. We focus on two
different kind of masses in this work, $40\,M_{\odot}$ and $10\,M_{\odot}$ (but
obviously the spectrum will not have only two components in nature). We find
that there can be as many as thousands (tens, in the $10\,M_{\odot}$ case) when
they are in their very early phase (i.e. when they are the farthest from merger
but with a SNR of at least 10), hundreds (a few, in the $10\,M_{\odot}$ case)
later, when they have lost a significant amount of their eccentricity, and a
few (basically zero for the $10\,M_{\odot}$ case) when they have circularised
\footnote{Visit one of these URLs for an illustration:\\ \href{https://youtu.be/zLJ6i6TR1Nk}{https://youtu.be/zLJ6i6TR1Nk} \\or\\ \href{https://tinyurl.com/bilibili-E-EMRIs}{https://tinyurl.com/bilibili-E-EMRIs}}.

Considering only continuous sources, we estimate the incoherent sum of the
individual contributions of E-EMRIs and find that their signal will reach
values as high as $10^{-16}$ in characteristic amplitude and as far as $3\times
10^{-2}$ Hz in frequency.  However, because of the computational
time, we have used a binning in frequency which is larger than we expect to
have with LISA. Indeed, depending on the mission duration, the binning will
allow us to discern some individual sources, which would have interesting
implications due to the large SNRs, as explained before.

This might pose a problem for polychromatic EMRIs, ``the ones we have been talking
about all along'' in the literature (to use Bernard Schutz' words), because they
might be partially or even completely buried in the signal created by E-EMRIs. The
problem becomes even harder when considering the white dwarf confusion signal.

As a final note, the sources we have addressed when calculating the total
signal have at least a full period for this treatment, i.e. we have considered
sources with $P_{\rm orb} \leq T_{\rm obs}$ because otherwise the Fourier
treatment would be ill-defined. The rest of the sources will contribute in a
different manner. Very eccentric sources only emit radiation when they are
close to pericentre i.e. over a short time span $dt_{\rm peri}$ which is very
short compared to the orbital period. In principle one could take $dt_{\rm
peri} \sim R_{\rm peri}/V_{\rm peri}$, where $R_{\rm peri}$ is the periapsis
distance and $V_{\rm peri}$ the associated velocity.  The probability that such
a source passes pericentre while LISA is operating is then simply $P_{\rm
orb}/T_{\rm obs}$ independently of $dt_{\rm peri}$ as long as $dt_{\rm peri}
\ll T_{\rm obs}$, which is the condition for the source to be considered as a
short burst, a ``pop''. We dub this kind of sources ``popcorn EMRIs'' and note
that even though they are not continuous signals, one can still Fourier
transform them using the family of all smooth (i.e. infinitely differentiable)
functions $f$ on $\mathbf{R}$ that vanish faster than any polynomial, the
Schwarz space. This will be presented elsewhere in a separate work.

\section*{Acknowledgments}

We thank Matteo Bonetti for the discussions about the incoherent sum of
amplitudes.  We acknowledge the funds from the ``European Union
NextGenerationEU/PRTR'', Programa de Planes Complementarios I+D+I (ref.
ASFAE/2022/014).

\end{document}